\documentclass{article}

\usepackage{arxiv}

\usepackage[utf8]{inputenc} 
\usepackage[T1]{fontenc}    
\usepackage{hyperref}       
\usepackage{url}            
\usepackage{booktabs}       
\usepackage{amsfonts}       
\usepackage{nicefrac}       
\usepackage{microtype}      
\usepackage{lipsum}		
\usepackage{graphicx}
\usepackage{doi}
\usepackage{multirow}
\usepackage{flafter}

\title{Design and Development of a Roaming Wireless Safety Emergency Stop}

\author{ \href{https://orcid.org/0000-0002-5390-3946}{\includegraphics[scale=0.06]{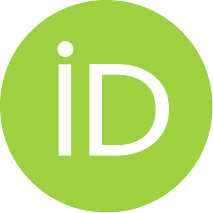}\hspace{1mm}Henry Beuster}\\
	Electrical Measurement Engineering\\
	Helmut-Schmidt-University\\
	Hamburg, Germany\\
	\texttt{henry.beuster@hsu-hh.de} \\
	\And
	\href{https://orcid.org/0000-0002-6882-1214}{\includegraphics[scale=0.06]{orcid.pdf}\hspace{1mm}Thomas Doebbert}\\
	Electrical Measurement Engineering\\
	Helmut-Schmidt-University\\
	Hamburg, Germany\\
	\texttt{thomas.doebbert@hsu-hh.de} \\
    \And
	\href{https://orcid.org/0000-0001-7490-6487}{\includegraphics[scale=0.06]{orcid.pdf}\hspace{1mm}Christoph Cammin}\\
	Institute of Automation Technology\\
	Helmut-Schmidt-University\\
	Hamburg, Germany\\
	\texttt{christoph.cammin@hsu-hh.de} \\
    \And
	{\hspace{1mm}Dmytro Krush}\\
	Electrical Measurement Engineering\\
	Helmut-Schmidt-University\\
	Hamburg, Germany\\
	\texttt{dmytro.krush@googlemail.com} \\
    \And
	{\hspace{1mm}Gerd Scholl}\\
	Electrical Measurement Engineering\\
	Helmut-Schmidt-University\\
	Hamburg, Germany\\
	\texttt{gerd.scholl@hsu-hh.de} \\
}

\date{}

\renewcommand{\shorttitle}{Roaming Wireless Safety Emergency Stop}

\hypersetup{
pdftitle={Design and Development of a Roaming Wireless Safety Emergency Stop},
pdfsubject={functional safety, wireless, IO-Link Wireless, roaming},
pdfauthor={Henry Beuster},
pdfkeywords={functional safety, wireless, IO-Link Wireless, roaming},
}

\begin{document}
\maketitle

\begin{abstract}
	Modern manufacturing is characterized by a high degree of automation, with autonomous systems also frequently being used. In such environments human intervention in the event of malfunctions or maintenance becomes a rare but also necessary task. When human workers are no longer an integral part of the production process, but only intervene when necessary, e.g., in the case of unexpected machine behavior, appropriate safety solutions will become even more important. This work describes a wireless communication system enabling a flexible and safe emergency stop function for multiple automation cells. A portable emergency stop switch allows seamless transition between different wireless cells, ensuring functional safety. The communication protocol combines IO-Link Wireless features with the safety requirements already implemented in IO-Link Safety. Security requirements are fulfilled through encryption and authentication. The IO-Link Wireless roaming functionality is used to extend the system across several manufacturing cells. An experimental setup confirms the suitability of the system for various applications. The results demonstrate the effectiveness of the handover mechanism and evaluate the potential of the system to improve flexibility, availability and security in dynamic production environments. Future extensions could include the use of AI based evaluation of the radio signals for an intelligent cell handover.
\end{abstract}

\keywords{functional safety \and wireless \and IO-Link Wireless \and roaming}

\section{Introduction}
In all areas of industrial automation, the need for secure, reliable, fast and flexible wireless communication solutions is ubiquitous. In contrast to previous wired communication solutions, wireless solutions offer greater local flexibility and a reduction of installation effort with significant cost savings (e.g., \cite{scholl_wireless_2013}). In addition to the consumer standards such as IEEE 802.11-based WLAN, Bluetooth, and in some cases, Zigbee, a number of specific wireless standards for predominantly industrial (sensor/actuator) data communication have become established (e.g., \cite{scholl_wireless_2013,doebbert_study_2021,krush_coexistence_2021}). In the public funded project “Digital Sensor-2-Cloud Campus Platform” (DS2CCP) \cite{helmut-schmidt-university_ds2ccp_nodate}, in which IO-Link Wireless (IOLW) and 5G technology are implemented as future-oriented and cross-platform solutions, a reliable and robust wireless communication of sensors and actuators between the industrial shop floor and an edge cloud could be demonstrated \cite{doebbert_study_2021, cammin_concept_2023,doebbert_testbed_2022}. The project aims for a consistent and holistic conception of an industrial DS2CCP for applications in the field of automation technology. Previous studies on requirements regarding safety, security and timing to aim for safety-related and security-for-safety applications are given in \cite{doebbert_study_2021,doebbert_safety_2022} and the first IO-Link Wireless Safety (IOLWS) protocol concept evaluation is completed with a positive approval \cite{doebbert_contribution_2024,doebbert_contribution_2023}. Therefore, in this contribution, a further step is approached by adding more flexibility to IOLWS utilizing roaming within multiple automation cells.\\
This paper is organized as follows: After this general introduction to safe wireless communication, a focus is especially set on IOLWS in Section \ref{sec:Sec2}. Section \ref{sec:Sec3} describes the roaming scenario and the measurement setup. The results are presented at the end of the section. The graphical evaluation and subsequent discussion follow in Section \ref{sec:Sec4}. Section \ref{sec:Sec5} summarizes the results and suggests further research opportunities.

\section{IO-Link Wireless Safety}\label{sec:Sec2}
\begin{figure}[tb]
	\centering
	\includegraphics[width=0.9\linewidth]{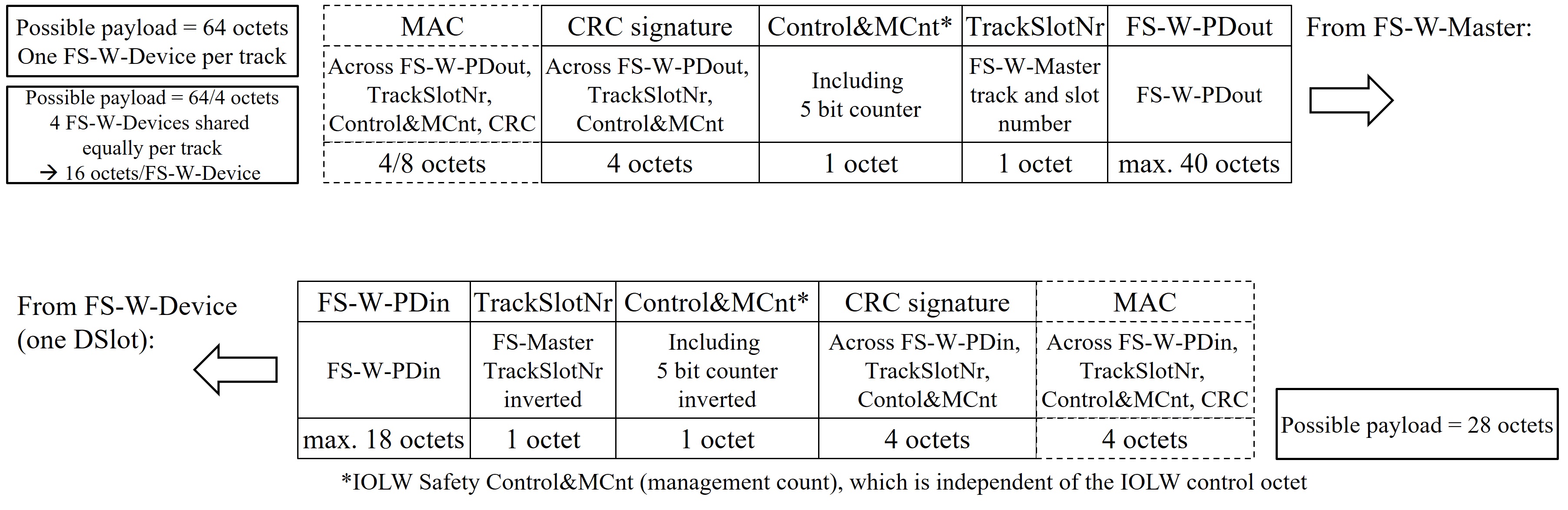}
	\caption{IOLWS Safety PDU based on \cite{doebbert_contribution_2024,doebbert_contribution_2023}.}
	\label{fig:fig1}
\end{figure}

The functional safe wireless communication protocol in \cite{doebbert_contribution_2024} combines the system architecture and basic communication algorithm of IOLW \cite{noauthor_iec_2023} and the safety requirements and principles of IO-Link Safety (IOLS) \cite{noauthor_iec_2022}. Recognizing the ongoing discussion and the inherent security challenges in wireless communication systems, especially in safety-critical applications, the proposed solution integrates encryption algorithms and authentication in an underlying security layer to follow security-for-safety requirements with a bit error rate up to 0.5, according to \cite{doebbert_contribution_2024,schiller_enhancement_2022}.\\
In line with IEC 61784-3 \cite{iec_61784-3_iec_2021} the security and safety communication layers within IOLWS are embedded into the software structure and the communication protocol. The Safety Process Data Unit with the concept of explicit transmission of the safety measures for timeliness (Control\&MCnt), authenticity (track/slot number), integrity as well as the point-to-point connection guaranteed by applying a cryptographic algorithm with sufficient entropy \cite{doebbert_contribution_2024} is illustrated in Figure \ref{fig:fig1}. Here, a Message Authentication Code (MAC) is employed to detect manipulations (guarantees integrity and authentication of the message) and a Cyclic Redundancy Check (CRC) to detect (stochastic) errors (data integrity) \cite{doebbert_contribution_2024}.\\
Within the configuration, the communication of non-safety wireless process data from the Failsafe-(FS-)W-Master using one communication channel per FS-W-Device is also possible. In this case, the functional safety wireless process data output length is limited to 22 octets. For the input data, it is also feasible to separate the six octets of the functional safety wireless process data input and equivalently non-safety wireless process data, which, therefore, are not included in the calculation of the MAC and CRC.

\section{Roaming with IO-Link Wireless Safety}\label{sec:Sec3}
\begin{figure}[tb]
	\centering
	\includegraphics[width=0.9\linewidth]{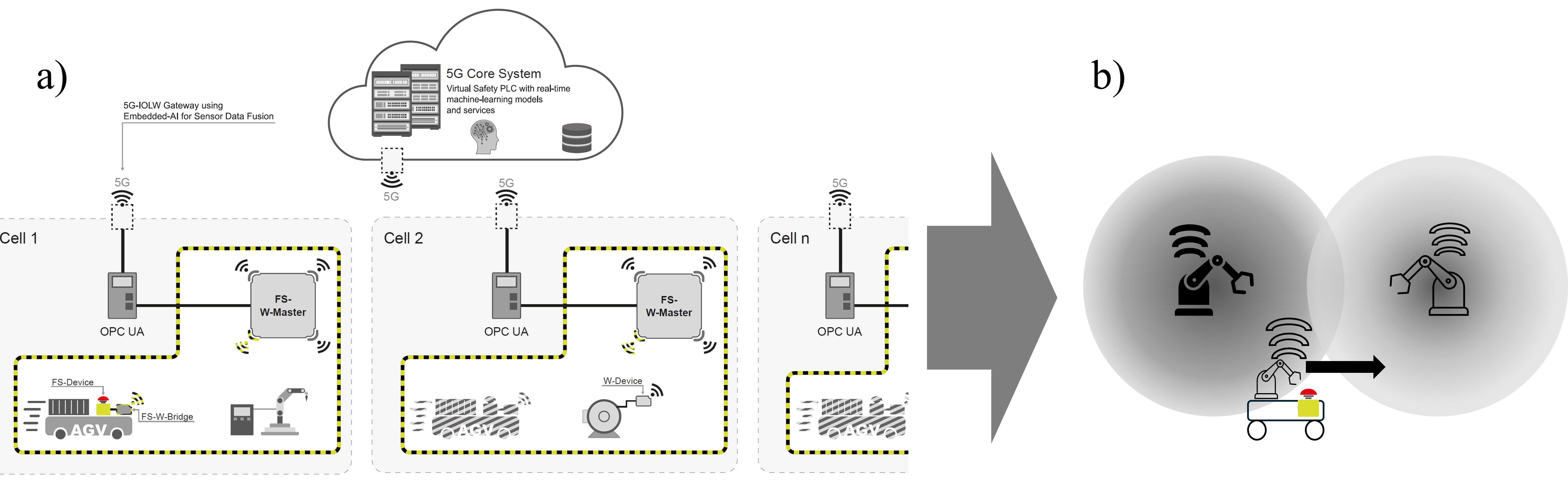}
	\caption{a) Modular sensor-2-cloud automation topology, based on \cite{doebbert_testbed_2022}. b) Simplified scenario comparison of two wireless cells and one roaming (FS-)W-Device.}
	\label{fig:fig2}
\end{figure}

A study on roaming, particularly in the context of IOLW, is given in \cite{rentschler_roaming_2017}, whereby the first published version of the IOLW standard already included the roaming feature. In this context, roaming describes the possibility that (FS-)W-Devices may connect to another cell (another (FS-)W-Master), when leaving the cell on purpose or disconnecting for any reason from the previous (FS-)W-Master \cite{rentschler_roaming_2017}.\\
In a simplified/reduced safety-related scenario, a mobile (FS-)W-Device (i.e., an emergency stop on a robot) moves from one cell to another, whereby the signal strength decreases at the boundary of the cell that is left and increases in the cell that is entered, as indicated by the grey scales in Figure \ref{fig:fig2} b). The measurement setup emulates the scenario depicted in Figure \ref{fig:fig2}.

\subsection{Measurement Setup}
\begin{figure}[tb]
	\centering
	\includegraphics[width=0.9\linewidth]{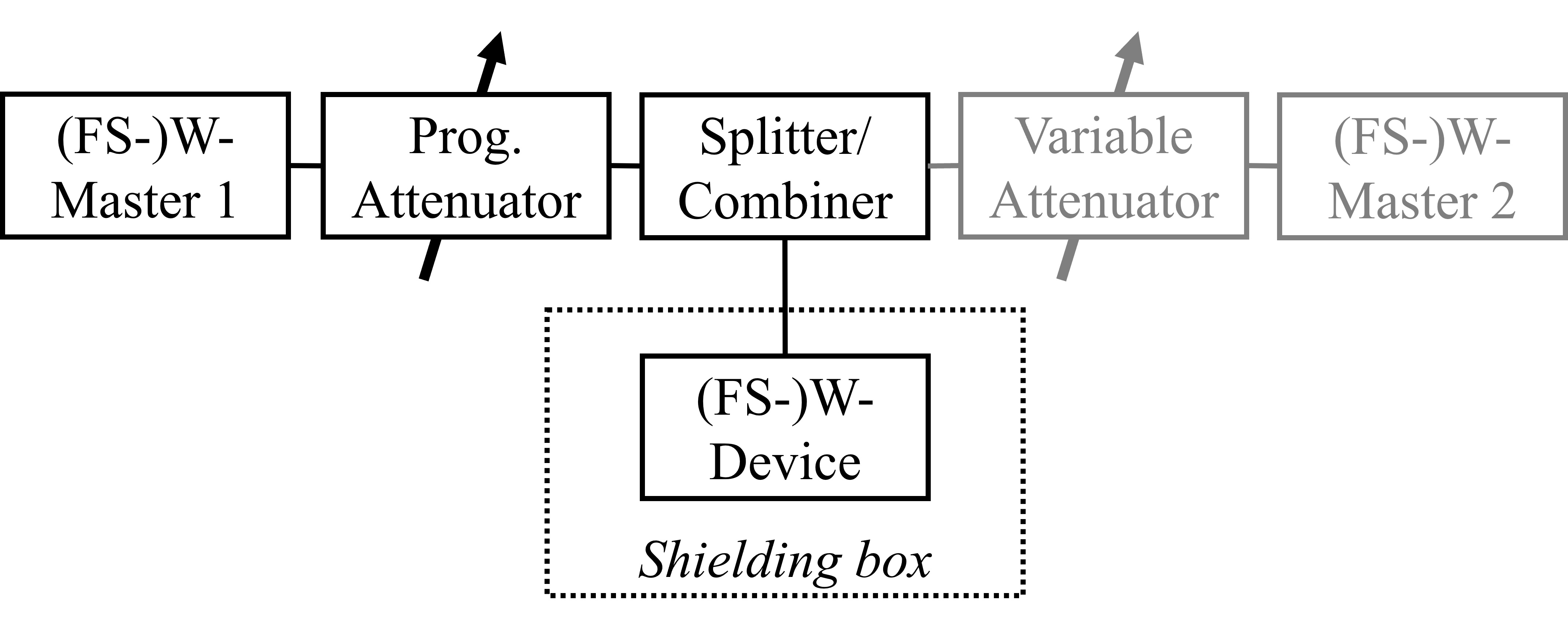}
	\caption{Measurement setup, comprising of two (FS-)W-Masters and one (FS-)W-Device.}
	\label{fig:fig3}
\end{figure}

Figure \ref{fig:fig3} illustrates a simplified setup for evaluation. In a wired shielded and reproducible setup, two (FS-)W-Masters are connected to a single (FS-)W-Device using a (resistive, i.e., non-directional) power splitter/combiner and variable/programmable attenuator. Controlling the attenuation of ''(FS-)W-Master\,2'', and thus adjusting the receiving signal power of the (FS-)W-Device, the scenario of movement between the cells (as sketched in Figure \ref{fig:fig2} b) is emulated.\\
Utilizing output pins for status observations connected to an oscilloscope, the transition time for the cell handover is measured and recorded. For a better overview, the computer for controlling the entire setup and the oscilloscope for time measurement are not shown here.

In total four measurement series are performed, two with a standard IOLW protocol stack and two with an IOLWS protocol stack. Two of them are carried out as ''preliminary'' measurements with only one (FS-)W-Master, as indicated by the (FS-)W-Master 1 box surrounded by the black solid line in Figure \ref{fig:fig3}. The remaining two measurement series are performed with both (FS-)W-Master\,1 and (FS-)W-Master\,2 (grey box).

\subsection{IO-Link Wireless Roaming Connection/Reconnection}
In the first measurement series, the duration of establishing the connection between a roaming W-Device and a W-Master in track mode ''Roaming'' and port mode ''RoamingAutoPairing'' is evaluated as preparatory study without the safety stack. W-Master and W-Device are set to maximum transmission power level, corresponding to 10\,dBm for the W-Master and 4\,dBm for the W-Device. The programmable attenuator is switched between ''ON'' (an adjustable attenuation to emulate different distances) and ''OFF'' (103\,dB, maximum attenuation), with a duration of two seconds each, i.e. four seconds in total for one measurement cycle. In the ''ON''-state attenuations are chosen to emulate the following connection conditions:
\begin{itemize}
	\item strong signal (30\,dB attenuation resulting in Received Signal Strength Indicator (RSSI) of -37\,dBm on the W-Master side, below any saturation effects),
	\item moderate connections (50 and 65\,dB attenuation, RSSI of -53 and -67\,dBm),
	\item weak connections (80 and 83\,dB attenuation, RSSI of -83 and -87\,dBm),
    \item bad connection (85\,dB attenuation, RSSI of -89\,dBm, just enough for a successful pairing).
\end{itemize}
Meanwhile, the duration for a reconnection is measured. In total 300 valid reconnections per setting are recorded with a sampling rate of 10,000 per second. The results for IOLW are given in Table \ref{tab:table1}. The mean IOLW roaming connect duration is always below 0.5\,s for attenuations smaller than 80\,dB, rapidly increasing beyond this limit.

\subsection{IO-Link Wireless Safety Roaming Connection/Reconnection}
Similar to the previous testcase, the procedure is repeated with the safety stack extension on both the FS-W-Master and the FS-W-Device. The results for IOLWS are given in Table \ref{tab:table1}.

\begin{table}[tb]
\caption{Results for IOLW and IOLWS roaming connection times.}
\begin{tabular}{|l|l|ll|ll|ll|ll|}
\hline
\multicolumn{1}{|c|}{\multirow{2}{*}{\textbf{\begin{tabular}[c]{@{}c@{}}Attenuation\\ “ON” {[}dB{]}\end{tabular}}}} & \multicolumn{1}{c|}{\multirow{2}{*}{\textbf{RSSI {[}dBm{]}}}} & \multicolumn{2}{c|}{\textbf{\begin{tabular}[c]{@{}c@{}}Minimum\\ duration {[}s{]}\end{tabular}}} & \multicolumn{2}{c|}{\textbf{\begin{tabular}[c]{@{}c@{}}Maximum\\ duration {[}s{]}\end{tabular}}} & \multicolumn{2}{c|}{\textbf{Mean duration {[}s{]}}} & \multicolumn{2}{c|}{\textbf{\begin{tabular}[c]{@{}c@{}}Standard\\ deviation {[}s{]}\end{tabular}}} \\ \cline{3-10} 
\multicolumn{1}{|c|}{} & \multicolumn{1}{c|}{} & \multicolumn{1}{c|}{\textbf{IOLW}} & \multicolumn{1}{c|}{\textbf{IOLWS}} & \multicolumn{1}{c|}{\textbf{IOLW}} & \multicolumn{1}{c|}{\textbf{IOLWS}} & \multicolumn{1}{c|}{\textbf{IOLW}} & \multicolumn{1}{c|}{\textbf{IOLWS}} & \multicolumn{1}{c|}{\textbf{IOLW}} & \multicolumn{1}{c|}{\textbf{IOLWS}} \\ \hline 
30 & -37 & \multicolumn{1}{l|}{0.429} & 0.454 & \multicolumn{1}{l|}{0.487} & 0.512 & \multicolumn{1}{l|}{0.450} & 0.475 & \multicolumn{1}{l|}{0.015} & 0.015 \\ \hline
50 & -53 & \multicolumn{1}{l|}{0.429} & 0.454 & \multicolumn{1}{l|}{0.487} & 0.512 & \multicolumn{1}{l|}{0.452} & 0.477 & \multicolumn{1}{l|}{0.016} & 0.016 \\ \hline
65 & -67 & \multicolumn{1}{l|}{0.429} & 0.454 & \multicolumn{1}{l|}{0.486} & 0.512 & \multicolumn{1}{l|}{0.455} & 0.480 & \multicolumn{1}{l|}{0.017} & 0.017 \\ \hline
80 & -83 & \multicolumn{1}{l|}{0.429} & 0.454 & \multicolumn{1}{l|}{1.132} & 1.157 & \multicolumn{1}{l|}{0.479} & 0.504 & \multicolumn{1}{l|}{0.075} & 0.075 \\ \hline
83 & -87 & \multicolumn{1}{l|}{0.457} & 0.482 & \multicolumn{1}{l|}{2.080} & 2.105 & \multicolumn{1}{l|}{0.812} & 0.837 & \multicolumn{1}{l|}{0.244} & 0.244 \\ \hline
85 & -89 & \multicolumn{1}{l|}{0.438} & 0.463 & \multicolumn{1}{l|}{5.913} & 5.938 & \multicolumn{1}{l|}{2.883} & 2.935 & \multicolumn{1}{l|}{1.390} & 1.410 \\ \hline
\end{tabular}
\label{tab:table1}
\end{table}

In comparison to the IOLW roaming connect testcase with the standard IOLW stack the results measured with the safety extension show an increased connection time of approx. 25\,ms in the mean due to an additional parameter exchange phase establishing the safety connection. This difference is less dependent on the attenuation than the duration of roaming connect itself because the safety layer connection is established via process data exchange of IOLW. This means that transmission errors are detected and avoided with a residual error probability of 10$^{-9}$ \cite{noauthor_iec_2023}, while these mechanisms are not yet effective in establishing the IOLW connection.

\subsection{IO-Link Wireless Handover}
For the measurement of the IOLW handover, the setup is extended with a second W-Master, as shown in Figure \ref{fig:fig3}. Also, the procedure is changed to emulate the intended use-case with a handover coordinated by the higher-level controller \cite{rentschler_roaming_2017}. The programmable and variable attenuator are set to the same attenuation for each series of measurements. A script running on the measuring computer controls the simultaneous and alternating pairing and unpairing via the Standardized Master Interface (SMI) of both W-Masters. The duration between loss of communication of one W-Master and the establishment of the connection to the other W-Master is measured 300 times, again with 10,000 samples per second, for each attenuation setting. A series with an attenuation higher than 80\,dB respectively an RSSI lower than -83\,dBm is not included for this setup, as the handover would not be successful in a reasonable time. In turn, a series of measurements with 77\,dB attenuation, RSSI of -80\,dBm, was carried out to better analyze the behavior in the limit range of reliable handover tests. Table \ref{tab:table2} shows the results of this testcase. The difference to the first two testcases results from the coexistence or shared use of the IOLW configuration channels by both W-Masters, which are needed to establish the connection.

\subsection{IO-Link Wireless Safety Handover}
The setup is the same for the last testcase, but with the safety stack used on both FS-W-Masters and the FS-W-Device. The results are given in Table \ref{tab:table2}. Also in this case a nearly constant difference of approx. 25\,ms between the non-safety and the safety mode due to the additional communication load can be observed.

\begin{table}[tb]
\caption{Results for IOLW and IOLWS handover.}
\begin{tabular}{|l|l|ll|ll|ll|ll|}
\hline
\multicolumn{1}{|c|}{\multirow{2}{*}{\textbf{\begin{tabular}[c]{@{}c@{}}Attenuation\\ “ON” {[}dB{]}\end{tabular}}}} & \multicolumn{1}{c|}{\multirow{2}{*}{\textbf{RSSI {[}dBm{]}}}} & \multicolumn{2}{c|}{\textbf{\begin{tabular}[c]{@{}c@{}}Minimum\\ duration {[}s{]}\end{tabular}}} & \multicolumn{2}{c|}{\textbf{\begin{tabular}[c]{@{}c@{}}Maximum\\ duration {[}s{]}\end{tabular}}} & \multicolumn{2}{c|}{\textbf{Mean duration {[}s{]}}} & \multicolumn{2}{c|}{\textbf{\begin{tabular}[c]{@{}c@{}}Standard\\ deviation {[}s{]}\end{tabular}}} \\ \cline{3-10} 
\multicolumn{1}{|c|}{} & \multicolumn{1}{c|}{} & \multicolumn{1}{c|}{\textbf{IOLW}} & \multicolumn{1}{c|}{\textbf{IOLWS}} & \multicolumn{1}{c|}{\textbf{IOLW}} & \multicolumn{1}{c|}{\textbf{IOLWS}} & \multicolumn{1}{c|}{\textbf{IOLW}} & \multicolumn{1}{c|}{\textbf{IOLWS}} & \multicolumn{1}{c|}{\textbf{IOLW}} & \multicolumn{1}{c|}{\textbf{IOLWS}} \\ \hline
30 & -37 & \multicolumn{1}{l|}{0.429} & 0.454 & \multicolumn{1}{l|}{0.487} & 0.512 & \multicolumn{1}{l|}{0.450} & 0.475 & \multicolumn{1}{l|}{0.015} & 0.015 \\ \hline
50 & -53 & \multicolumn{1}{l|}{0.429} & 0.454 & \multicolumn{1}{l|}{0.487} & 0.512 & \multicolumn{1}{l|}{0.452} & 0.477 & \multicolumn{1}{l|}{0.016} & 0.016 \\ \hline
65 & -67 & \multicolumn{1}{l|}{0.429} & 0.454 & \multicolumn{1}{l|}{0.486} & 0.512 & \multicolumn{1}{l|}{0.455} & 0.480 & \multicolumn{1}{l|}{0.017} & 0.017 \\ \hline
80 & -83 & \multicolumn{1}{l|}{0.429} & 0.454 & \multicolumn{1}{l|}{1.132} & 1.157 & \multicolumn{1}{l|}{0.479} & 0.504 & \multicolumn{1}{l|}{0.075} & 0.075 \\ \hline
83 & -87 & \multicolumn{1}{l|}{0.457} & 0.482 & \multicolumn{1}{l|}{2.080} & 2.105 & \multicolumn{1}{l|}{0.812} & 0.837 & \multicolumn{1}{l|}{0.244} & 0.244 \\ \hline
85 & -89 & \multicolumn{1}{l|}{0.438} & 0.463 & \multicolumn{1}{l|}{5.913} & 5.938 & \multicolumn{1}{l|}{2.883} & 2.935 & \multicolumn{1}{l|}{1.390} & 1.410 \\ \hline
\end{tabular}
\label{tab:table2}
\end{table}

\section{Evaluation}\label{sec:Sec4}
\begin{figure}[htb]
	\centering
	\includegraphics[width=0.9\linewidth]{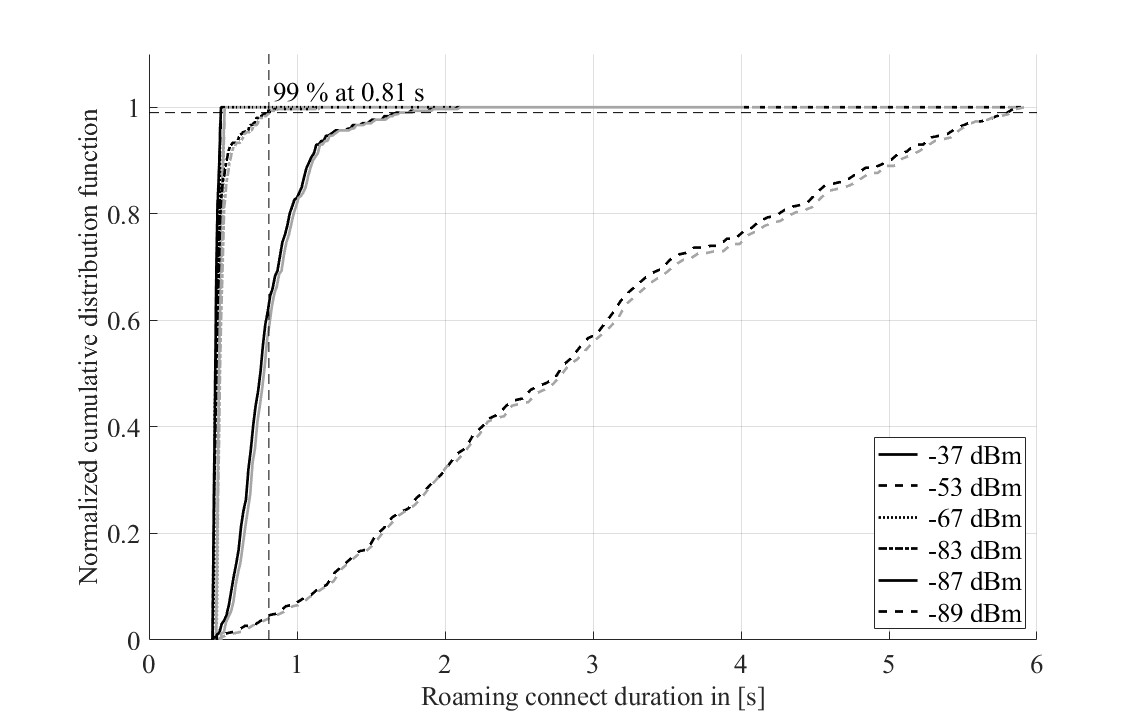}
	\caption{Duration for roaming connect in non-safety (black) and safety (grey) mode for the given RSSI (flatter lines represent lower RSSIs).}
	\label{fig:fig4}
\end{figure}

The graphical evaluation for the one-sided roaming (re-)connection durations is given in Figure \ref{fig:fig4}. Each line represents the connect duration for a given RSSI and safety mode as empirical Cumulative Distribution Function (eCDF). All graphs approach to one as the attenuation is chosen for successful pairing. The eCDFs for strong and moderate RSSIs are very similar and close to each other, showing that 100\,\% of the connections are established under 487\,ms for non-safety and under 512\,ms in safety mode. Pairing is also successful under poorer connection conditions, but more likely to take longer than 500\,ms or even several seconds. A reference point is given at 99\,\% successful connections faster than 0.81\,s at an RSSI of -83\,dBm to mark a suitable measurement below one second.

\begin{figure}[htb]
	\centering
	\includegraphics[width=0.9\linewidth]{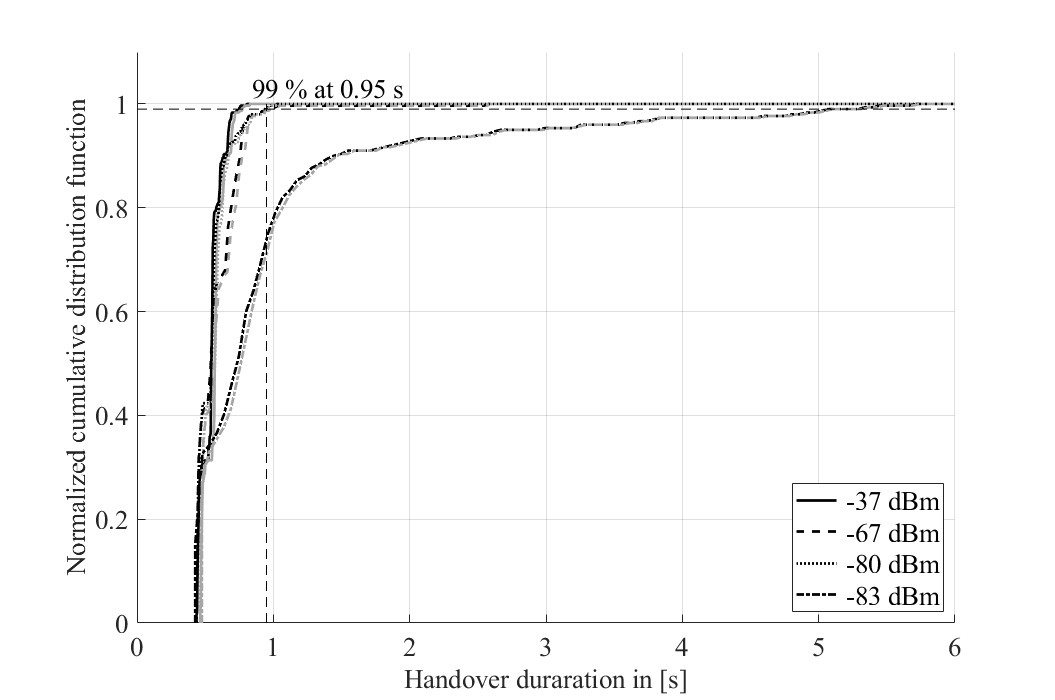}
	\caption{Duration for handover in non-safety (black) and safety (grey) mode for the given RSSI.}
	\label{fig:fig5}
\end{figure}

In Figure \ref{fig:fig5} the graphical illustration of Table \ref{tab:table2} is carried out the same way. The graphs show the eCDFs of the measured durations for the handover from one (FS-)W-Master to the other in IOLW (non-safety mode) and IOLWS (safety mode) with the given RSSI. In these cases, the mean duration for good and moderate conditions is slightly increased by approx. 100\,ms with some outliers up to approx. 2.6\,s for the moderate condition. In contrast to establishing a single roaming connection to establish the handover connection, the original connection must be properly closed in the first step. The (FS-)W-Device is then unpaired and paired again with the other (FS-)W-Master. This causes additional delay. Another reason is the shared use of the configuration channels by both (FS-)W-Masters simultaneously. The reference point is given as 0.95\,s for 99\,\% successful handovers at an RSSI of -80\,dBm. The average handover time nearly doubles with a decrease of the RSSI value for a stable connection by -3\,dB, indicating that an RSSI of less than -80\,dBm is insufficient for many applications.

\section{Conclusion and Outlook}\label{sec:Sec5}
Under weak link conditions with up to an RSSI of -80\,dBm, safe link establishment and handover can be achieved in less than one second with a success rate of over 99\,\%. These conditions are equivalent to the free-space path loss at a distance exceeding 67\,m, while IOLW is rated for a maximum range of 10\,m in industrial environments with multiple active (FS-)W-Masters. Thus, there is enough system margin available. Using IOLWS instead of IOLW adds a nearly constant offset of 25\,ms and thus enables the usage of a functional safe wireless communication protocol in every use case IOLW roaming is suitable, e.g., in flexible and modular productions cells with a quick safety communication check after each line change is sufficient to re-arm the safety function. In case of a safety function response time greater than the handover duration, even a seamless handover could be possible with a sufficient signal strength. To meet safety requirements in a more dynamic and fast-moving use case further precautions are necessary. With additional location information, provided by, e.g., electronically readable tags, light curtains or an intelligent location system, the handover can possibly be triggered at a safe moment, in a safe area or by an intelligent algorithm with movement prediction. A radio channel analyzer combined with live AI based evaluation is applicable to optimize the channel configuration of the FS-W-Masters, e.g., if not all roaming cells are needed. In an advanced and demanding application in terms of availability and timing it is also conceivable to equip one FS-W-Device with two radios to realize a seamless handover.

\section*{Acknowledgement}
This research is funded by dtec.bw – Digitalization and Technology Research Center of the Bundeswehr. dtec.bw is funded by the European Union – NextGenerationEU (project “Digital Sensor-2-Cloud Campus Platform” (DS2CCP) with the project website \cite{helmut-schmidt-university_ds2ccp_nodate}).

\bibliographystyle{unsrt}
\bibliography{references}

\end{document}